\documentclass[prb,twocolumn,showpacs,superscriptaddress]{revtex4}
\usepackage{graphicx}

\begin{document}

\title{Transport properties of strongly correlated electrons in quantum dots 
using a simple circuit model}

\author{G. B. Martins}
\email[corresponding author: ]{martins@oakland.edu}
\affiliation{Department of Physics, Oakland University, Rochester, MI 48309}
\author{C. A. B\"usser}
\affiliation{Condensed Matter Sciences Division, Oak Ridge National Laboratory, Oak Ridge,
Tennessee 37831}
\affiliation{Department of Physics and Astronomy, The University of Tennessee, Knoxville,
Tennessee 37996}
\author{K. A. Al-Hassanieh}
\affiliation{Condensed Matter Sciences Division, Oak Ridge National Laboratory, Oak Ridge,
Tennessee 37831}
\affiliation{Department of Physics and Astronomy, The University of Tennessee, Knoxville,
Tennessee 37996}
\affiliation{National High Magnetic Field Laboratory and Department of Physics, Florida State
University, Tallahassee, FL 32306}
\author{E. V. Anda}
\affiliation{Departamento de F\'{\i}sica, Pontif\'{\i}cia Universidade Cat\'olica do Rio de Janeiro, 
22453-900, Brazil}
\author{A. Moreo}
\affiliation{Condensed Matter Sciences Division, Oak Ridge National Laboratory, Oak Ridge,
Tennessee 37831}
\affiliation{Department of Physics and Astronomy, The University of Tennessee, Knoxville,
Tennessee 37996}
\author{E. Dagotto}
\affiliation{Condensed Matter Sciences Division, Oak Ridge National Laboratory, Oak Ridge,
Tennessee 37831}
\affiliation{Department of Physics and Astronomy, The University of Tennessee, Knoxville,
Tennessee 37996}

\begin{abstract}
Numerical calculations are shown to reproduce the main results of recent 
experiments  involving nonlocal spin control in
nanostructures (N. J. Craig {\it et al.}, Science {\bf 304}, 565 (2004)). 
In particular, the splitting of the zero-bias-peak discovered
experimentally is clearly observed in our studies. 
To understand these results, a simple ``circuit model''
is introduced and shown to provide a good qualitative description of the experiments.
The main idea is that the splitting originates in a Fano anti-resonance, which 
is caused by having one quantum dot
side-connected  in relation to the current's path. This scenario provides
an explanation of Craig {\it et al.}'s results that is alternative to the
RKKY proposal, which is here also addressed.
\end{abstract}

\maketitle

The observation of the Kondo effect in a single quantum dot (QD) \cite{Goldhaber1} 
and the subsequent theoretical and experimental studies of more complex 
structures, such as two QDs directly 
coupled through a tunable potential barrier\cite{kouwen}, has provided 
impetus for the analysis of more elaborate systems. In a recent seminal work, 
Craig {\it et al.} \cite{craig} report on the possible laboratory 
realization of the two-impurity Kondo system. Two similar QDs are 
coupled through an open conducting central region (CR). A finite bias 
is applied to one of the QDs (QD1 from now on) as well as 
to the CR, while the other QD (QD2) 
is kept at constant gate potential. The differential conductance 
of QD1 is then measured for different charge states of QD2 and different 
values of its coupling to the CR. The main result was the suppression and 
splitting of the zero-bias-anomaly (ZBA) in QD1 by changing the occupancy of QD2 
from even to odd number of electrons and by increasing its coupling 
to the CR. A Ruderman-Kittel-Kasuya-Yosida (RKKY) interaction between 
the QDs was suggested as an explanation for the observed effects \cite{glazman}. 
The importance of Craig {\it et al.}'s experiments
cannot be overstated: the possibility of performing 
nonlocal spin control in a system with two lateral QDs has potential applications 
in QD-based quantum computing \cite{divincenzo}. 

In this Letter, numerical simulations 
in good agreement with the experiments are presented. 
The central conclusion of this work is that our computational data, and as a consequence
the experimental results, can be explained using 
a very simple ``circuit model'', where one of the elements is a T-connected
QD that has an intrinsic reduction of conductance with varying biases.
This proposal is an alternative to the more standard RKKY ideas. 
\begin{figure}[h]
\centering
\includegraphics[height=3.25cm]{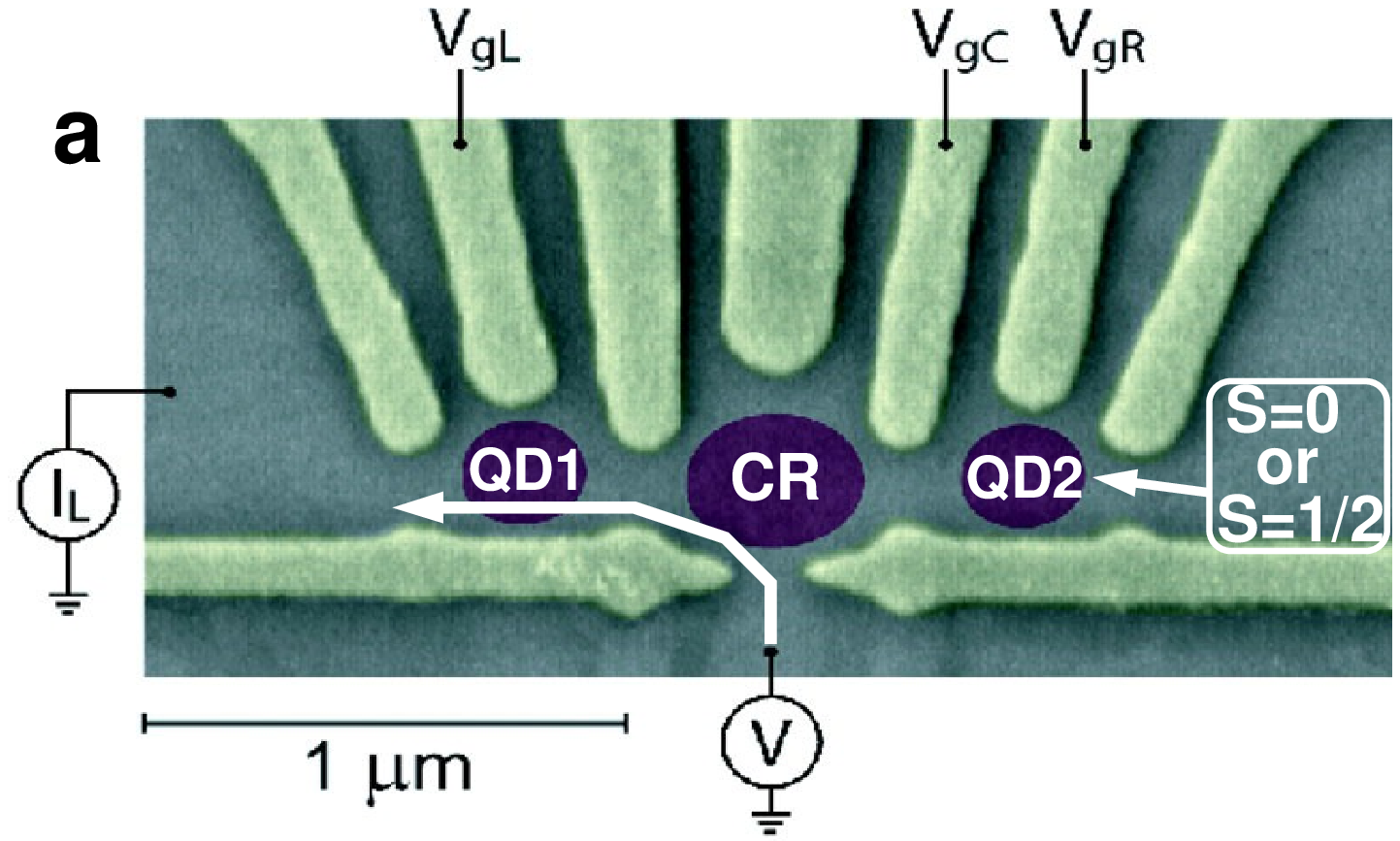}
\includegraphics[height=1.75cm]{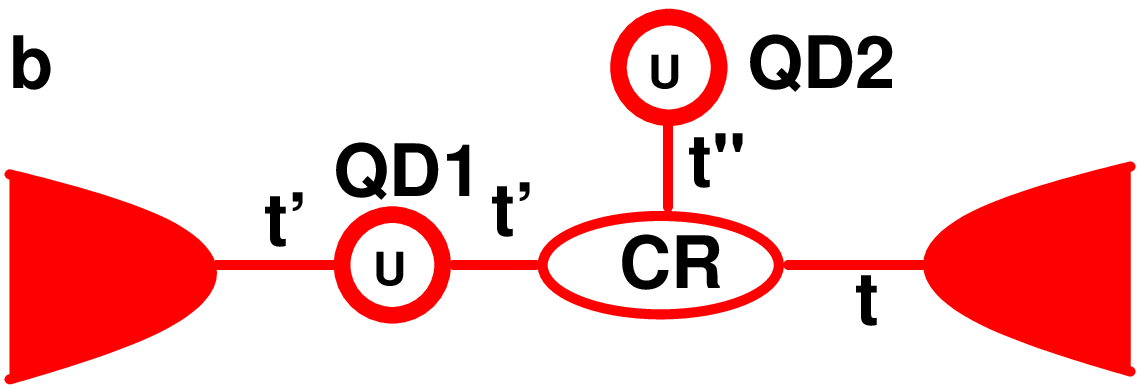}
\caption{ (a) Experimental setup used in Ref. \onlinecite{craig}. 
(b) Illustration of the model studied in this Letter (see text for details).}
\label{fig1}
\end{figure}
Fig. 1a depicts the experimental set up used in the measurements of 
Craig {\it et al.} \cite{craig} with the labeling used in this Letter.
Figure 1b is a schematic representation of the system, introducing two different 
tunneling parameters (hopping matrix elements $t^{\prime}$ and $t^{\prime \prime}$) 
and the Coulomb repulsion U in each QD (assumed the same for simplicity).
To model this system, the Anderson impurity Hamiltonian is used for both QDs:
\begin{flushright}
\begin{eqnarray}
H_{\rm d}=\sum_{i=1,2;\sigma} \left[ U n_{i \sigma} n_{i \bar{\sigma}} + 
V_{\rm gi} n_{i \sigma} \right] ,
\end{eqnarray}
\end{flushright}
where the first term represents the usual Coulomb repulsion between
two electrons in the same QD, 
and the second term is the effect of the gate potential $V_{\rm gi}$ over each QD. 
QD1 is directly connected to the left lead and to the CR 
with hopping amplitude $t'$, while 
QD2 is connected only to the CR (with hopping amplitude $t^{\prime \prime}$), 
which itself is
connected to the right lead with hopping amplitude $t$ (which is also the 
hopping amplitude in both leads, and our energy scale). In summary, 
\begin{eqnarray}
H_{\rm leads} &=& t \sum_{i \sigma} \left[ c_{l i\sigma}^{\dagger} c_{l i+1\sigma}
 +  c_{r i\sigma}^{\dagger} c_{r i+1\sigma} +\mbox{h.c.} \right], \\
H_{12} &=& \sum_{\sigma} \left[ t^{\prime} c_{1 \sigma}^{\dagger} \left(c_{l0 \sigma} 
+ c_{\rm CR \sigma} \right) + t^{\prime \prime}  
c_{2 \sigma}^{\dagger} c_{\rm CR \sigma} + \right. \nonumber \\ && 
\left. tc_{\rm CR \sigma}^{\dagger}c_{r0 \sigma}+\mbox{h.c.} \right],
\end{eqnarray}
where $c_{l i\sigma}^{\dagger}$ ($c_{r i\sigma}^{\dagger}$ ) creates an electron at
site $i$ with spin $\sigma$ in the left (right) lead. 
The CR is composed of one tight-binding site \cite{note2}, unless otherwise stated. Site `0' is
the first site at the left (right) of QD1 (CR) in the left (right) lead.
The total Hamiltonian is $H_{\rm T} = H_{\rm d} + H_{\rm leads} + H_{12}$.
Note that for $V_{\rm g1}=V_{\rm g2}=-U/2$, the Hamiltonian is particle-hole symmetric.
To calculate the conductance $G$, using the Keldysh formalism \cite{Meir-cnd}, 
a cluster containing the interacting dots and 
a few sites of the leads is solved exactly\cite{note1}, the Green 
functions are calculated, and the leads are 
incorporated through a Dyson Equation embedding procedure. Details of the 
embedding have been extensively discussed before\cite{interfere}. 
All the results shown were obtained for $U=0.5$, $t^{\prime}=0.2$, zero-bias, and zero temperature. 

\begin{figure}[h]
\includegraphics[height=5.8cm]{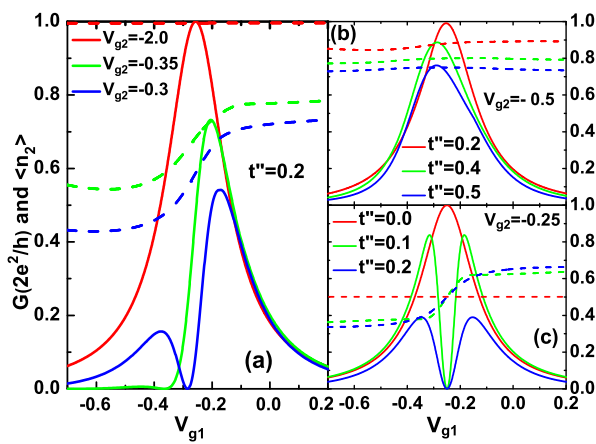}
\caption{
(a) Variation of $G$  with $V_{\rm g1}$ in QD1 (solid curves) and 
variation of $\langle n_2\rangle$ (occupancy of 
QD2 per spin orientation - dashed curves) for $t^{\prime \prime}=0.2$ 
and three different values of $V_{\rm g2}$. 
For $V_{\rm g2}=-2.0$ (red), QD2 is occupied by 2 electrons ($\langle n_2\rangle=1$)  for any
$V_{\rm g1}$ and the conductance through QD1 
is essentially the same as if QD2 was not present. For higher values of $V_{\rm g2}$, the average 
value of $\langle n_2\rangle$ decreases and becomes dependent on $V_{\rm g1}$ 
(decreasing for lower values of $V_{\rm g1}$). This
is accompanied by a suppression of the ZBA (for $-0.35$ (green)) 
and also by a splitting of the ZBA (for $V_{\rm g2}=-0.3$ (blue)).
(b) Variation of $G$ and $\langle n_2\rangle$ with 
$t^{\prime \prime}$ (0.2, 0.4, 0.5) at a fixed value of $V_{\rm g2}=-0.5$. As the value of 
$t^{\prime \prime}$ increases, the average value of $\langle n_2\rangle$ decreases 
and this is again accompanied by a suppression of the ZBA. 
(c) Same as in (b), but now for $V_{\rm g2}=-0.25$ (particle-hole symmetric point) 
and $t^{\prime \prime} = 0.0$, $0.1$, and $0.2$. 
Note that $G$ vanishes at $V_{\rm g1}=-0.25$, where $\langle n_2\rangle=0.5$, for all finite values 
of $t^{\prime \prime}$.
}
\label{fig2}
\end{figure}

In Fig. 2, results for the conductance across QD1 (solid curves) and for the occupancy 
per spin orientation $\langle n_2\rangle$ 
of QD2 (dashed curves) are presented. 
In Fig. 2a, $t^{\prime \prime}=0.2$ and $V_{\rm g2}$ varies from $-2.0$ to $-0.3$.
For $V_{\rm g2}=-2.0$ (red), QD2 is occupied by two electrons ($\langle n_2\rangle=1$) and the 
conductance of QD1 displays the characteristic Kondo behavior reported
before \cite{comment1}.
For $V_{\rm g2}=-0.35$ (green) the average value of $\langle n_2\rangle$ decreases to $\approx 0.7$ 
($\approx 1.4$ electrons in QD2) and $\langle n_2\rangle$ now depends
on $V_{\rm g1}$. In addition, $G$ decreases in comparison to the result obtained 
for $V_{\rm g2}=-2.0$. 
Then, these numerical results are  
qualitatively in agreement with the experimental results 
shown in Fig. 2 of Craig {\it et al.} \cite{craig}, 
namely, by decreasing the occupancy of QD2, from even to odd number of 
electrons, the ZBA in QD1 is suppressed.
As $V_{\rm g2}$ is further increased ($-0.3$ (blue)) a {\it qualitative} change occurs:
For values of $V_{\rm g1}$ where $\langle n_2\rangle \approx 0.5$ (QD2 singly occupied), the conductance of 
QD1 vanishes and therefore there is a narrow dip in $G$.  This splitting of the ZBA 
is remarkably similar to that observed in Fig. 3A of the experimental results\cite{craig}. 
For finite-temperature calculations, 
the dip in $G$ will not reach zero, resembling even better the experiments\cite{anda}. 

To further test the similarities between simulations and experiments, 
in Figs. 2b and 2c results for $G$ and $\langle n_2\rangle$ are shown for fixed $V_{\rm g2}$  
and different $t^{\prime \prime}$ values. 
In Fig. 2b, where $V_{\rm g2}=-0.5$, as $t^{\prime \prime}$ increases from $0.2$ to $0.5$
there is only a slight decrease of $G$. This is accompanied by a 
slight decrease in the average value of $\langle n_2\rangle$, from $\approx 0.9$ to $\approx 0.7$. 
A more dramatic change is obtained in Fig. 2c, where $V_{\rm g2}=-0.25$, and $t^{\prime \prime}$ varies 
from 0.0 to 0.2. By increasing $t^{\prime \prime}$ from $0.0$ (red curves) to 
$0.1$ (green), the ZBA is now split in two and $\langle n_2\rangle$ 
acquires a dependence on $V_{\rm g1}$.
As $t^{\prime \prime}$ further increases ($0.2$ (blue)), the dip becomes wider, the two side-peaks decrease 
and $G$ still vanishes for $\langle n_2\rangle=0.5$ (one electron in QD2).
Our calculations show that, if $\langle n_2\rangle$ varies around $0.5$, the dip in $G$ is present for 
all finite values of $t^{\prime \prime}$, with a width proportional to $t^{\prime \prime}$.
Comparing the results in Figs. 3A and 3B of Craig {\it et al.} 
\cite{craig} with Figs. 2c and 2b in this Letter, respectively, one notices a striking 
similarity: The splitting of the ZBA observed in the experimental results 
(their Fig. 3A), when the number of electrons in the control QD is odd and the coupling to the central 
region is increased, is very similar to the dip in $G$ for all 
finite-$t^{\prime \prime}$ curves in Fig. 2c (as mentioned above, at 
finite temperatures, one expects that the dip in $G$ will not reach zero). 
When the occupancy of QD2 is even (Fig. 3B in the experimental results \cite{craig} and Fig. 
2b in this Letter), the $G$ dependence on $t^{\prime \prime}$ is much 
less significant and the splitting of the ZBA does not occur.

\begin{figure}[h]
\centering
\includegraphics[width=4.8cm]{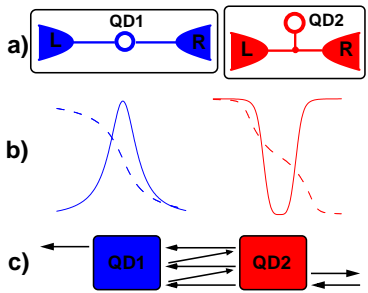}
\caption{ Schematic representation of the main ideas behind the ``circuit model''. 
In (a), the system represented in Fig. 1b is divided into its constituent 
elements: QD1 is modeled as a QD connected in series with the leads and QD2 is 
modeled as a side-connected QD. The curves in (b) represent the 
conductance and occupancy of each separate circuit element vs. the
applied gate potential. (c) Schematic representation of how the 
two individual elements are connected to form the final `circuit': Incident 
and reflected wave amplitudes are represented in the right side of QD2 by black arrows. 
A transmitted wave through QD2 undergoes multiple reflections between the two quantum dots 
until it is finally transmitted past QD1. The superposition of all these processes results 
in the final conductance for the `circuit'.
}
\label{fig5}
\end{figure}

What is the origin of these results?
Below, it will be argued that a qualitative description of the results can be achieved 
by analyzing the two quantum dots through a so-called `circuit model'. 
This model starts
with the conductance of each QD calculated separately, as independent elements 
of a circuit, and then the conductance of the `complete circuit' is obtained by 
combining the conductances of the two elements connected in series. 
Fig. 3 describes schematically the steps involved in this approach. In Fig. 3a, 
the complete system formed by QD1 and QD2 (shown in Fig. 1b) 
is divided into two components. QD1 is modeled as 
a QD connected directly to left (L) and right (R) leads, while QD2 
is modeled as a side-connected QD\cite{sato}. Fig. 3b shows the respective conductances 
and occupancies for each independent element vs. gate voltage,
and Fig. 3c represents the 
scattering processes (represented by transmission and reflection amplitudes) 
that an electron undergoes while moving through the complete `circuit'. 
The superposition of all these processes leads to the total
transmittance (proportional 
to the conductance)  for the circuit model. This can be calculated in two ways: 
coherently or incoherently\cite{data}. 
Since there is no qualitative difference between them, and in order to 
keep the simplicity of the model, we present the incoherent results. 
The equation which provides 
the final transmittance for the processes depicted in Fig. 3c is 
\begin{eqnarray}
T=\frac{T_1T_2}{1-R_1R_2},
\end{eqnarray}
where the transmittances $T_1$ and $T_2$ are proportional to the conductances for QD1 and QD2, 
as depicted in Fig. 3b, and $R_{1(2)}=1-T_{1(2)}$ are the reflectances. 
To calculate $T$, one needs to establish how $T_2$ depends on $V_{\rm g1}$. The natural 
way to do that is to use the dependence of $\langle n_2\rangle$ on $V_{\rm g1}$, as depicted 
in Fig. 2, and then use the relation between conductance and occupancy, 
as shown in the red curves in Fig. 3b. In other words, the functional relation 
can be expressed as $T_2=T_2(\langle n_2\rangle (V_{\rm g1}))$. It is not surprising that in a strongly 
correlated system like the one being analyzed here, the variation of the gate potential 
of QD1 will influence the charge occupancy of QD2, and in turn this will influence 
the conductance through QD1.

\begin{figure}[h]
\includegraphics[height=5.3cm]{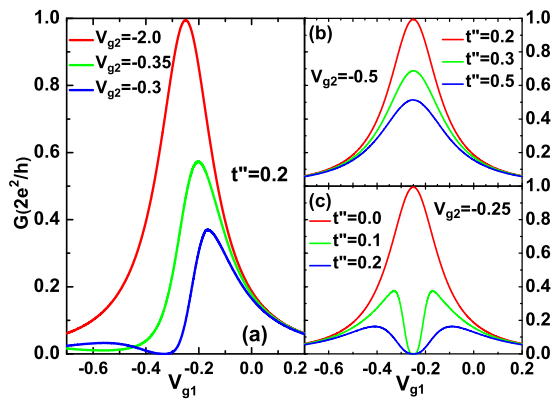}
\caption{ Same as in Fig. 2, but now using the `circuit model' for the 
calculations.
}
\label{fig6}
\end{figure} 

In Fig. 4, conductance results using Eq.(4) are shown for the same parameters 
as in Fig. 2. Although the quantitative agreement varies, there is 
good overall qualitative agreement. All the trends are correctly reproduced 
and some of the details are quite similar, such as for example the asymmetric shape 
of the curves at higher values of $V_{\rm g2}$ ($-0.35$ and $-0.3$) in Fig. 4a. 
It is important to notice that there are {\it no} adjustable parameters in 
the circuit model here presented. The only input necessary is $\langle n_2 \rangle$ 
vs. $V_{\rm g1}$, which is obtained through a calculation 
for the complete system (values displayed for $\langle n_2\rangle$ in Fig. 2). 
The success of the circuit model implies that the dip in $G$ arises from the 
Fano anti-resonance which cancels the conductance of QD2 (red solid curve in Fig. 3b). 
The Fano anti-resonance can be seen as a destructive interference process between two
different trajectories an electron can take on its way to QD1: it can cross the CR 
without passing through QD2; or it can visit QD2, return to the CR and then proceed 
to QD1 \cite{sato}.

The similarities between the experimental results and our simulations suggest
that our model and numerical technique have captured the essential physics 
of the experiments. However, these same experiments have also been explained
using RKKY ideas\cite{glazman}. Can our numerical results be also understood
in this alternative context?
To try to answer 
this question, several calculations were performed with 
different parameter values and number of sites in the CR \cite{new}. 
In Fig. 5a, results for spin correlations between QD1 and QD2 (denoted 
${\bf S_1 \cdot S_2}$) are presented for the same parameters used in Fig. 2c. 
At $t^{\prime \prime}=0.0$ (red curve) QD1 and QD2 are uncorrelated
as expected. 
As $t^{\prime \prime}$ increases to $0.1$
(green), and then $0.2$ (blue), it is observed that in the region
where $G$ reaches its maximum value (see Fig. 2c), ${\bf S_1 \cdot S_2}$ also assumes a
maximum value and it is positive (ferromagnetic (FM)).
For $t^{\prime \prime}>0.2$ (not shown), ${\bf S_1 \cdot S_2}$ saturates and starts 
decreasing.
The maximum of ${\bf S_1 \cdot S_2}$, for all values of $t^{\prime \prime}$, 
decreases even further as the size of the central region increases (the results in Fig. 5a 
are for a CR with just one site). In addition, the sign of ${\bf S_1 \cdot S_2}$ alternates as the 
size of the CR increases and the QDs move farther apart from each other. 
In Fig. 5b, results for the spin correlation between QD1 and
its neighboring site in the CR (denoted ${\bf S_1 \cdot S_c}$) is shown for the 
same parameters as in Fig. 5a. ${\bf S_1 \cdot S_c}$ is a rough measure
of the Kondo correlation in QD1, having a direct connection with the ZBA 
in Fig. 2c. Indeed, for $t^{\prime \prime}=0.0$ (red) when
$G$ reaches the unitary limit, a robust antiferromagnetic (AF)
correlation develops between QD1 and its neighboring site in the CR. 
For $t^{\prime \prime}=0.1$ (green), 
despite the narrow dip in $G$, the side-peaks are still close to the unitary limit
(see Fig. 2c) and ${\bf S_1 \cdot S_c}$ is still strongly AF. However, for 
$t^{\prime \prime}=0.2$ (blue),  
both $G$ and ${\bf S_1 \cdot S_c}$ are strongly suppressed, in qualitative
agreement with a suppressed ZBA due to a weakened Kondo resonance.

\begin{figure}[h]
\includegraphics[width=7.55cm]{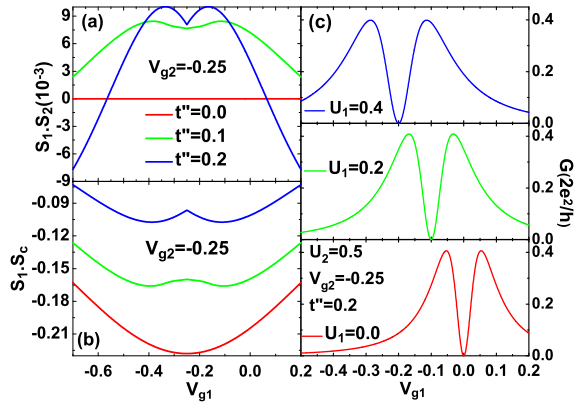}
\caption{ (a) Spin correlation ${\bf S_1 \cdot S_2}$ between QD1 and QD2 for the same parameters 
as in Fig. 2c. For $t^{\prime \prime}=0.0$ (red curve), the two QDs are uncorrelated
(${\bf S_1 \cdot S_2}=0$). For finite $t^{\prime \prime}$ ($0.1$ (green) and $0.2$ (blue)), 
${\bf S_1 \cdot S_2}$ is FM
and reaches its maximum value in the region where $G$ is maximum. 
(b) Kondo correlations ${\bf S_1 \cdot S_c}$ between QD1 and the central site for the same
parameters as in (a). All values are AF and they decrease in amplitude as 
$t^{\prime \prime}$ increases, underscoring the decrease of the Kondo effect as the FM correlation 
between QD1 and QD2 increases (compare with (a)). 
(c) Variation of $G$ as $U_1$ (Hubbard interaction in QD1) assumes the values 
$0.4$, $0.2$ and $0.0$. Note that the dip in $G$ becomes slightly narrower as $U_1$ decreases, 
however it does not disappear.
}
\label{fig3}
\end{figure}

The results thus far
seem to indicate that the CR could be mediating a long range coupling between
QD1 and QD2, with the characteristics of an RKKY interaction. However, the 
magnitude of the maximum value of ${\bf S_1 \cdot S_2}$ (see scale
in Figs. 5a-b) is too small 
to account for all the effects observed in the conductance in Fig. 2c. 
One possible way of increasing ${\bf S_1 \cdot S_2}$ is by 
coupling QD1 more strongly to the CR than to the left lead. This was exactly 
the setup chosen in Ref. \onlinecite{craig}, where those authors 
performed the measurements with asymmetric couplings to the left ($\Gamma_{\rm L}$) and right 
($\Gamma_{\rm CR}$) sides of QD1. 
In fact, the voltages applied to the gates in Fig. 1a 
were such that $\Gamma_{\rm CR} \gg \Gamma_{\rm L}$. 
In our model, this is equivalent to having an asymmetric $t^{\prime}$, 
with $t^{\prime}_{\rm CR} \gg t^{\prime}_{\rm L}$. An analysis of the results 
in this asymmetric regime indicates that the correlation between QD1 and QD2 does indeed 
increase. However, if one performs the calculations with the sites in the CR at a filling 
lower than one electron per site (half-filling), it is observed 
that ${\bf S_1 \cdot S_2}$ is 
gradually suppressed as the electron filling falls to a more appropriate level to 
simulate the two-dimensional electron gas in the CR. 
Although one can argue that some of 
the dependence of the conductance of QD1 on the charge state of QD2 seen in Fig. 2 is 
associated to the correlations between the two dots, it is apparent that other effects are 
also present.  This is dramatically exemplified by the fact that
the cancellation of $G$ presented
in Fig. 2c occurs for {\it any} 
finite value of $t^{\prime \prime}$, and of course 
for $t^{\prime \prime} \approx 0$, 
one finds that ${\bf S_1 \cdot S_2} \approx 0$. 
The fact that the dip seen in the 
conductance in Fig. 2c is not dominantly caused by 
correlations between the dots can be made more clear 
by checking the results for the conductance as $U_1$ (Hubbard interaction 
in QD1) is reduced to zero. In Fig. 5c, results for $G$ are shown 
for 3 different values of $U_1$, for the same parameters as for the blue curve in Fig. 2c. 
As $U_1$ decreases from $0.4$ (blue) to $0.2$ (green), and then to $0.0$ (red), the dip in the 
conductance remains, only becoming narrower, indicating that its origin is not associated 
with many-body interactions, but more likely with cancellations typical
of T-geometries \cite{sato} that occur even in the non-interacting limit.

In summary, the numerical results qualitatively reproduce
the main aspects of important recent experiments \cite{craig} involving 
nonlocal spin control in nanostructures. 
The main result is that the splitting observed in the ZBA is 
caused by a cancellation in the conductance due
to a destructive interference. This so-called Fano anti-resonance has its origin 
in one of the dots being side-connected to the current's path. 
A simple `circuit model' qualitatively reproduces the experiments and offers 
an alternative to a purely RKKY interpretation of the results, underscoring 
that a laboratory realization of the two-impurity Kondo system should avoid 
any geometry susceptible to a Fano anti-resonance.

C.B., K. A., A. M., and E. D. are supported by NSF grant DMR 0454504. G. M. by an 
OU internal grant.


\begin{thebibliography}{}

\bibitem{Goldhaber1} D. Goldhaber-Gordon {\it et al.}, Nature {\bf 391}, 156 (1998).

\bibitem{kouwen} W. G. van der Wiel {\it et al.} Rev. Mod. Phys. {\bf 75}, 1 (2003).

\bibitem{craig} N. J. Craig {\it et al.}, Science {\bf 304}, 565 (2004).

\bibitem{glazman} L. I. Glazman {\it et al.}, Science {\bf 304}, 524 (2004); 
P. Simon {\it et al.}, Phys. Rev. Lett. {\bf 94}, 086602 (2005); 
M. G. Vavilov {\it et al.}, Phys. Rev. Lett. {\bf 94}, 086805 (2005).

\bibitem{divincenzo} D. Loss {\it et al.}, Phys. Rev. A {\bf 57}, 120 (1998).

\bibitem{note2} Calculations for a CR with a larger number of sites did not present 
any qualitatively new results.

\bibitem{Meir-cnd} Y. Meir {\it et al.}, Phys. Rev. Lett. {\bf 66}, 3048 (1991).

\bibitem{note1} Results shown include one site on each side. 
Calculations with more sites were done, but no significative size-effects were observed.

\bibitem{interfere} See V. Ferrari {\it et al.} Phys. Rev. Lett. {\bf 82}, 5088 (1999).

\bibitem{comment1} Removing QD2 results in essentially the 
same conductance as the one for $V_{\rm g2}=-2.0$. 

\bibitem{anda} R. Franco {\it et al.}, Phys. Rev. B {\bf 67}, 155301 (2003).

\bibitem{sato} M. Sato {\it et al.}, cond-mat/0410062.

\bibitem{data} S. Datta, {\it Electronic Transport in Mesoscopic Systems}, 
Cambridge University Press (1995). 

\bibitem{new} A full discussion of results will be presented in
G. B. Martins {\it et al.}, in preparation.

\end{thebibliography}
\end{document}